\documentclass[preprint, 12pt]{elsarticle}

\usepackage{amssymb}
\usepackage{amsmath}
\usepackage[breaklinks=true,colorlinks,citecolor=blue,linkcolor=blue,urlcolor=blue]{hyperref}

\journal{Journal of Magnetism and Magnetic Materials}

\begin{document}

\begin{frontmatter}

%% Title, authors and addresses

%% use the tnoteref command within \title for footnotes;
%% use the tnotetext command for theassociated footnote;
%% use the fnref command within \author or \affiliation for footnotes;
%% use the fntext command for theassociated footnote;
%% use the corref command within \author for corresponding author footnotes;
%% use the cortext command for theassociated footnote;
%% use the ead command for the email address,
%% and the form \ead[url] for the home page:
%% \title{Title\tnoteref{label1}}
%% \tnotetext[label1]{}
%% \author{Name\corref{cor1}\fnref{label2}}
%% \ead{email address}
%% \ead[url]{home page}
%% \fntext[label2]{}
%% \cortext[cor1]{}
%% \affiliation{organization={},
%%             addressline={},
%%             city={},
%%             postcode={},
%%             state={},
%%             country={}}
%% \fntext[label3]{}

\title{Coherent coupling between YBCO superconducting resonators and sub-micrometer-thick YIG films}

%% use optional labels to link authors explicitly to addresses:

 \author[label1]{Alberto Ghirri}
 \ead{alberto.ghirri@nano.cnr.it}
 \author[label2]{Mattia Cavani}
 \author[label2,label1]{Claudio Bonizzoni}    
 \author[label2,label1]{Marco Affronte}
 
 \affiliation[label1]{organization={Istituto Nanoscienze - CNR, Centro S3},
             addressline={via G. Campi 213/A},
             city={Modena},
             postcode={41125},
             country={Italy}}

 \affiliation[label2]{organization={Dipartimento di Scienze Fisiche, Informatiche e Matematiche, Universita di Modena e Reggio Emilia},
             addressline={via G. Campi 213/A},
             city={Modena},
             postcode={41125},
             country={Italy}}

%% Abstract
\begin{abstract}
%% Text of abstract
In cavity magnonics, magnon-photon hybridization has been widely investigated for both fundamental studies and applications. Planar superconducting resonators operating at microwave frequencies have demonstrated the possibility to achieve high couplings with magnons by exploiting the confinement of the microwave field in a reduced volume. Here we report a study of the coupling of high-$T_c$ YBCO superconducting waveguides with 104-nm-thick YIG magnetic films. We study the evolution of mode frequencies as a function of temperature and extract the coupling strength of hybrid magnon-photon modes. We show that the experimental results can be reproduced using a simple model in which the temperature dependence of the penetration depth accounts for the evolution of the polaritonic spectrum.

\end{abstract}

%%Graphical abstract
%\begin{graphicalabstract}
%\includegraphics{grabs}
%\end{graphicalabstract}

%%Research highlights
%\begin{highlights}
%\item Research highlight 1
%\item Research highlight 2
%\end{highlights}

%% Keywords
\begin{keyword}
YIG \sep YBCO  \sep Magnons \sep Microwaves \sep Strong coupling \sep Hybrid systems \sep Superconducting resonators

%% keywords here, in the form: keyword \sep keyword

%% PACS codes here, in the form: \PACS code \sep code

%% MSC codes here, in the form: \MSC code \sep code
%% or \MSC[2008] code \sep code (2000 is the default)

\end{keyword}

\end{frontmatter}

%% Add \usepackage{lineno} before \begin{document} and uncomment 
%% following line to enable line numbers
%% \linenumbers

%% main text
%%

\section{Introduction}
\label{sec:introduction}

Magnon-photon hybrids are obtained by coherently coupling collective excitations in coupled spin systems, that is spin waves and their quantized counterpart magnons, with microwave photon modes. These systems have proven to be fertile ground for both fundamental studies and applications \cite{RameshtiPhysRep22}. In the former case, topics like quantum magnonics \cite{Lachance-QuirionScieAdv17, YuanPhysRep22}, nonlinear effects \cite{WangPRL18, LeePRL23, BiPRL24}, ultrastrong coupling \cite{BourhillPRB16, GolovchanskiyPRAppl21, GhirriPRAppl23}, non-Hermitian physics \cite{ZhangNatCommun17, YuPhysRep24}, Floquet engineering \cite{XuPRL20}, magnonic frequency combs \cite{RaoPRL23} have been recently investigated. Applications in magnonics and quantum technologies, either at room temperature or in the quantum regime at mK temperatures, are related, but not restricted, to computation, sensing, microwave-to-optical transduction, coherent coupling of remote physical systems, and emission of microwaves, among others \cite{Lachance-QuirionApplPhysExp19, LiJAP20, PirroNatRevMater21, RameshtiPhysRep22, ChumakIEEE22}. 

For these studies, magnetic materials such as electrically insulating Yttrium Iron Garnet (YIG) or conducting permalloy, which are characterized by high spin densities and low ferromagnetic resonance (FMR) damping rates, have found widespread attention \cite{HanAPL24}. Incorporation of these materials into superconducting circuits has opened additional possibilities \cite{DobrovolskiyNatPhys19, GolovchanskiyAdvScie19, BorstScience23, BottcherNatPhys24}. Microwave modes in planar superconducting resonators have been exploited to achieve strong \cite{HueblPRL13, MorrisScieRep17, HouPRL19, LiPRL19} and ultrastrong \cite{GolovchanskiyJAP18, GolovchanskiyScAdv21, GolovchanskiyPRAppl23, GhirriPRAppl23, GhirriPRAppl24} coupling with magnons, also using nanomagnets fabricated directly onto the resonator \cite{HouPRL19, LiPRL19}. Furthermore, superconducting qubits were coupled to YIG spheres to detect single-magnon excitations \cite{Lachance-QuirionScieAdv17, XuPRL23}.

High critical temperature ($T_c$) superconducting resonators, in particular YBa$_2$Cu$_3$O$_7$ (YBCO), show $T_c$ up to about 90~K and resilience in applied magnetic field \cite{GhirriAPL15}. Strong and ultrastrong coupling with spin systems have been achieved, respectively, using paramagnetic ensembles \cite{BonizzoniAdvPhysX18, BonizzoniNPJ20, BonizzoniApplMagnReson23, Velluire-PellatScieRep23, BonizzoniNPJQuant24} and few-$\mu\mathrm{m}$-thick YIG films \cite{GhirriPRAppl23, GhirriPRAppl24}. In the latter case, the evolution of polaritonic modes at temperature below $T_c$ has shown a progressive frequency shift of the hybrid magnon-photon mode, which has been correlated with the penetration depth in the superconducting layer as an effect of the interplay between Meissner currents and spin waves \cite{GhirriPRAppl24}.

%%%
\begin{figure}[t]
\centering
\includegraphics[width=0.8\linewidth]{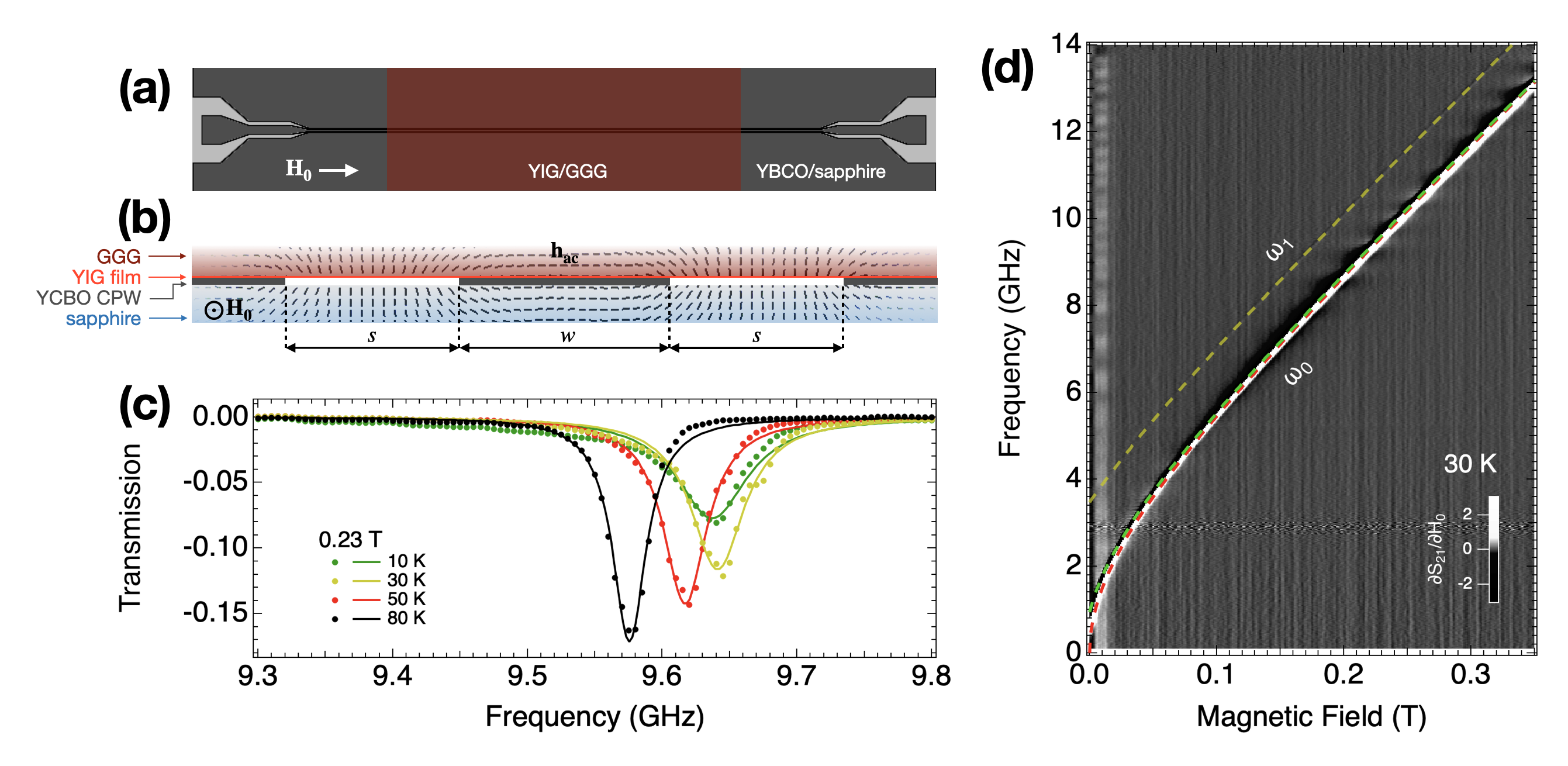}
\caption{Top view (a) and vertical section (b) of the CPW broadband line with the YIG film placed in top. Dashes in (b) show the distribution of the oscillating magnetic field $\mathbf{h_{ac}}$. (c) Evolution, at different temperatures, of the transmission spectra measured at $\mu_0H_0=0.23~\mathrm{T}$. The background $S_{21}$ spectrum obtained from off resonance data has been subtracted. Solid lines display the Lorentzian fit (Eq.~\ref{eq:Lorentzian}). (d) Spectral map showing the numerical derivative of broadband transmission spectra acquired at $T=30~\mathrm{K}$. Green and yellow dashed lines show $\omega_0$ and $\omega_1$ as derived respectively from Eq.~\ref{eq:KS_film_0} and Eq~\ref{eq:KS_film_1}. The red dashed line show $\omega_{FMR}$. }
\label{fig:broadband_1}
\end{figure}
%%%

The spectrum of spin-wave excitations in the YIG film and the coupling with the resonator depend upon the thickness of the magnetic layer. Here we experimentally investigate the evolution of transmission spectra obtained by placing a 104-$\mathrm{nm}$-thick YIG film both on a coplanar waveguide (CPW) broadband line and half-wavelength resonator fabricated from a YBCO film. We first study the broadband spectrum of excitations acquired using the YBCO CPW at different temperatures. We then analyze the coupling between microwave and magnon modes using the CPW resonator. We finally discuss how the extracted physical quantities compare with those obtained with thicker YIG films, which have been previously reported in Refs.~\cite{GhirriPRAppl23, GhirriPRAppl24}.  

\section{Experimental methods}
\label{sec:methods}

CPW broadband line and resonator were fabricated from a YBCO film having thickness of $330~\mathrm{nm}~(\pm 10 \%)$, which was deposited by reactive co-evaporation (RCE) \cite{KinderPhysicaC97} on a 3''-wide, 0.43-$\mathrm{mm}$-thick, r-cut sapphire substrate with a 20-nm-thick CeO$_2$ buffer layer (Ceraco GmbH, M-type). The characteristic porous surface of the YBCO film shows enhanced flux pinning and critical current density higher than $2 \times 10^6~\mathrm{A/cm^2}$ at 77~K. After deposition, the substrate was diced into $8 \times 5 \times 0.43~ \mathrm{mm^3}$ blocks, which were patterned using optical lithography and dry etching by Ar plasma in a Reactive Ion Etching (RIE) chamber. The central conductor of the CPW line has characteristic width $w=17~\mathrm{\mu m}$ and separation $s=14~\mathrm{\mu m}$ between the central conductor and the lateral ground planes (Fig.~\ref{fig:broadband_1}(a,b)). The broadband CPW supports transmission up to 14~GHz; the half-wavelength resonator has the same geometry as the CPW broadband line except for two 140-$\mathrm{\mu m}$-wide input and output capacitive gaps that interrupt the central conductor to define a central strip having length of 6~mm \cite{GhirriPRAppl23, GhirriPRAppl24}.

The YIG film that was grown by liquid phase epitaxy with thickness of 104~nm on a 0.5-mm-thick Gadolinium Gallium Garnet (GGG) substrate with (111) orientation (Matesy GmbH) \cite{KimPRMater25}. The reported FMR linewidth is less than 2~Oe at room temperature.

The YIG/GGG film was cut in a $\approx 4 \times 2~\mathrm{mm^2}$ sample and placed on the superconducting CPW. The film is gently pushed in contact with the YBCO surface by a plastic screw from the GGG side. The device was cooled down in a cryogenic setup having a superconducting solenoid that generates the external magnetic field ($H_0$), which is applied in the YIG film plane along the central conductor of the CPW (Fig.~\ref{fig:broadband_1}(a,b)). Transmission ($S_{21}$) measurements were carried out using a Vector Network Analyzer by sweeping the frequency at stationary values of the magnetic field. The incident microwave power is $\approx -8~\mathrm{dBm}$ at the CPW input port. Test experiments carried out in the same conditions using a bare GGG sample have not shown any detectable magnetic resonance lines. The derivatives of the transmission, $\partial S_{21}/\partial H_0$ and $\partial^2 S_{21}/\partial H_0^2$, were calculated numerically.

\section{Broadband transmission spectroscopy}
\label{sec:broadband}

We first used the broadband superconducting waveguide to study the evolution of ferromagnetic resonance lines. Transmission spectra evidence the presence of a main resonance at the frequency $\omega_0$ (Fig.~\ref{fig:broadband_1}(c,d)). As the temperature ($T$) decreases, $\omega_0$ progressively shifts toward higher frequencies. This trend indicates a decrease in the fixed-frequency resonance field as the temperature decreases between 80~K and 30~K. We fit the absorption dip, obtained by subtracting the background from the transmission spectra, using a Lorentzian curve \cite{PooleFarach79}
\begin{equation}
    \bar{S}_{21}=\frac{a}{1+\left(\frac{\omega-\omega_0}{\frac{1}{2}\Delta\omega}\right)^2},
    \label{eq:Lorentzian}
\end{equation}
where $\omega$ is the frequency, $a$ is the amplitude and $\Delta\omega$ is the full width at half maximum (Fig.~\ref{fig:broadband_1}(c)). We note that $\omega_0/2 \pi$ varies between 9.58~GHz at 80~K to 9.64~GHz at 10~K. Although the Lorentzian line shape does not perfectly reproduce the resonance lines in Fig.~\ref{fig:broadband_1}(c) we can obtain an estimate of $\Delta\omega/2\pi$, which increases from $30~\mathrm{MHz}$ at 80~K to $64~\mathrm{MHz}$ at 10~K. 

\begin{figure*}[t]
\centering
\includegraphics[width=\linewidth]{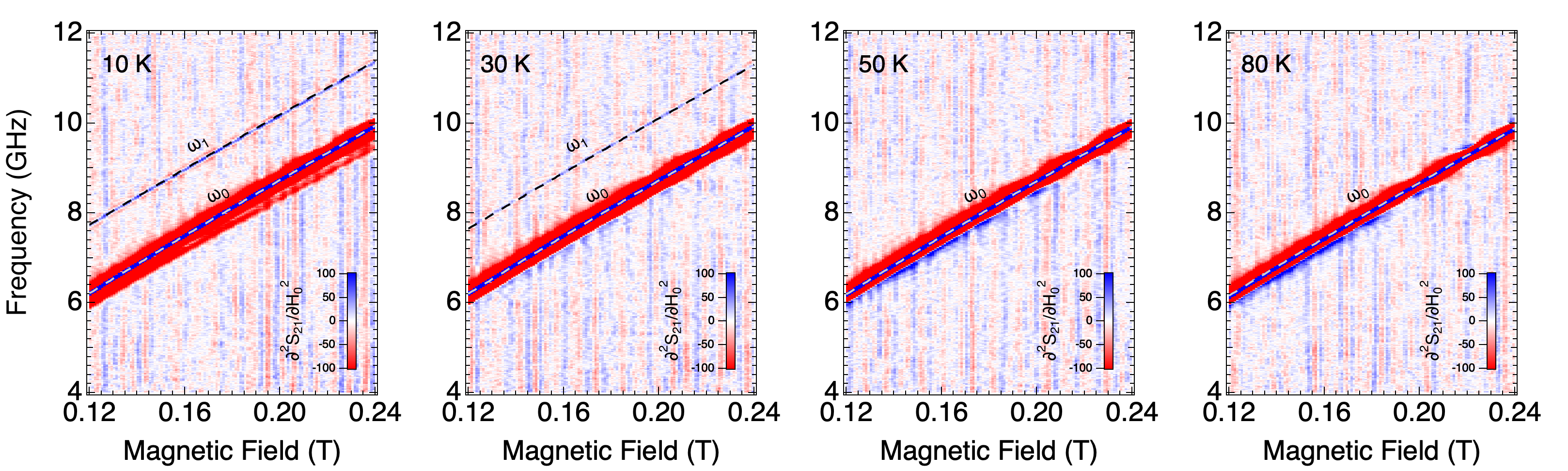}
\caption{Evolution of spectral maps measured at different temperatures with the broadband CPW. The color scale shows the second derivative of transmission with respect to the magnetic field. Dashed lines display the curves calculated with Eq.~\ref{eq:KS_film_0} and Eq.~\ref{eq:KS_film_1}, as indicated.}
\label{fig:broadband_2}
\end{figure*}

When the YIG and YBCO layers are in good contact, an additional faint line can be observed at low temperature, whose frequency $\omega_1/2\pi$, is approximately $1.2~\mathrm{GHz}$ above $\omega_0/2\pi$ (Fig.~\ref{fig:broadband_1}(d)). Fig.~\ref{fig:broadband_2} shows transmission maps taken at temperatures between 80 and 10~K, plotted as second derivative $\partial^2S_{21}/\partial H_0^2$ to evidence $\omega_0$ and $\omega_1$. We note that spectra taken above 50~K show the presence of $\omega_0$ only, while $\omega_1$ is visible at 30~K and below.

%%%
%%%

Considering the lateral profile of the broadband CPW and the Damon-Eshbach geometry in our experiment, we expect that the transmission line excites spin wave modes with wavenumber up to $k_y=3 \times 10^5~\mathrm{rad~m}^{-1}\approx 2\pi/s$ \cite{MaksymovPhysE15, GhirriPRAppl24}. According to the analytical model by Kalinikos ans Slavin (KS) \cite{KalinikosJPhysC86}, the frequency of the lowest mode can be calculated as \cite{DemokritovSpringer21}  
%%%%
\begin{align}
    \begin{split}
    & \omega_0 / 2 \pi= \\ & = \mu_0 \gamma \sqrt{H_0(H_0+M_s)+(M_s)^2 P_{00}(k_yd) \left(1-P_{00}(k_yd)\right)}, 
    \label{eq:KS_film_0}
    \end{split}
\end{align}
%%%%%
where $\gamma=28.02~\mathrm{GHz/T}$ is the electron gyromagnetic ratio and $P_{00}=1+[(1-\exp(-k_y d))/k_y d]$. We note that, being $d=104~\mathrm{nm}$ and $k_yd=0.03$, Eq.~\ref{eq:KS_film_0} results very close to $\omega_{FMR}=\mu_0 \gamma \sqrt{H_0(H_0+M_s)}$ (Fig.~\ref{fig:broadband_1}(d)). The temperature dependence of Eq.~\ref{eq:KS_film_0} derives from the saturation magnetization of YIG, $M_s$, which follows~\cite{SoltJAP62, MaierFlaigPRB17, GhirriPRAppl24} 
\begin{equation}
    M_s=M_0(1-u T^{3/2} - v T^{5/2}),
    \label{eq:Ms}
\end{equation}
being $u=23 \times 10^{-6}~\mathrm{K^{-3/2}}$ and $v=1.08 \times 10^{-7}~\mathrm{K^{-5/2}}$ \cite{MaierFlaigPRB17}. Using $\mu_0 M_0=0.28$~T, we can reproduce the magnetic field dependence of $\omega_0$ in the entire temperature range (Fig.~\ref{fig:broadband_1}(c,d) and~\ref{fig:broadband_2}).

Higher spin wave resonances beyond $\omega_0$ are described in the KS model as Perpendicular Standing Spin Wave (PSSW) modes \cite{KalinikosJPhysC86}. The frequency of the first PSSW mode approximately follows \cite{DemokritovSpringer21}
%%%%%
\begin{align}
    \begin{split}
        &\omega_1/2\pi = \\ & = \gamma \mu_0 \sqrt{\left[H_0+\frac{2A}{M_s}\left(k_y^2+\left(\frac{\pi}{d}\right)^2\right)\right]\left[H_0+\frac{2A}{M_s}\left(k_y^2+\left(\frac{\pi}{d}\right)^2\right) +M_s+H_0 \left(\frac{Ms/H_0}{\pi/d}\right)^2 k_y^2\right]},
    \label{eq:KS_film_1}
     \end{split}
\end{align}
%%%%%%%%
where $A$ is the exchange constant of the YIG film. The experimental spectra in Figs.~\ref{fig:broadband_1} and \ref{fig:broadband_2} can be reproduced using $A = 5.2~\mathrm{pJ/m}$ at 30~K and $A = 5.5~\mathrm{pJ/m}$ at 10~K.

\section{Coupling between magnetic film and resonator}
\label{sec:resonator}

The 104-nm-thick YIG film was subsequently placed on top of the superconducting resonator to test the coupling between magnon and photon modes. At low temperature, the bare YBCO/sapphire resonator shows the fundamental mode at $\omega_c/2 \pi\approx10~\mathrm{GHz}$, which decreases to $\omega_c/2\pi\approx 9.7~\mathrm{GHz}$ in the presence of the YIG/GGG sample.

\begin{figure*}[t]
\centering
\includegraphics[width=\linewidth]{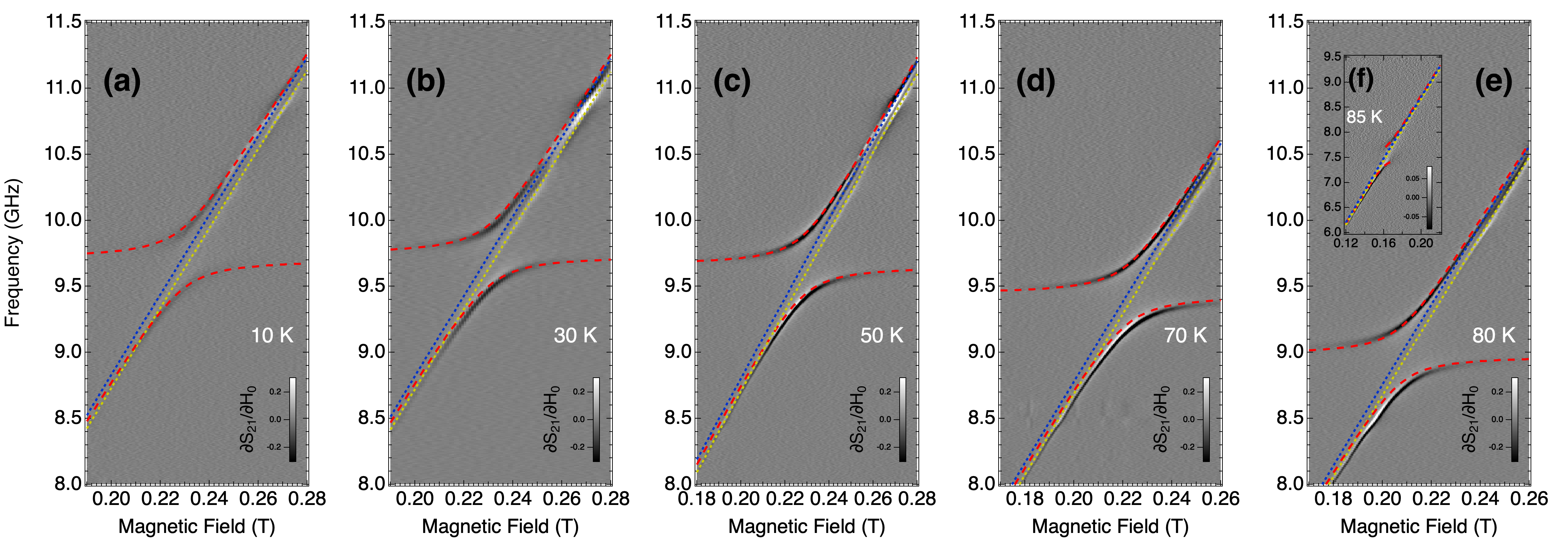}
\caption{Evolution of transmission ($\partial S_{21}/\partial H_0$) spectra acquired as a function of temperature using the CPW resonator}. The calculated polaritonic modes $\Omega_{\pm}$ are represented by red dashed lines; blue and yellow dashed lines show $\omega_0$ and $\omega_b$, respectively.
\label{fig:resonator_data}
\end{figure*}

Magnon-photon hybridization is shown by the formation of polariton branches in transmission spectra displayed for a series of temperatures in Fig.~\ref{fig:resonator_data}. Below 50~K (panels (a-c)), the splitting of approximately 460~MHz corresponds to the collective coupling $g/2\pi \approx 230~\mathrm{MHz}$. As the temperature increases, we observe the progressive lowering of the resonator frequency and the decrease in polariton splitting (panels (d-f)). Above $T_c$ the anticrossing disappears.

%%%
%%%

In order to extract meaningful parameters from the experimental spectra and compare them with previous results, we follow the analysis reported in \cite{GhirriPRAppl24}. The upper ($\Omega_+$) and lower ($\Omega_-$) polaritons can be described as the effect of the coupling between the resonator and a single magnetic mode ($\omega_b$). Their evolution follows \cite{GhirriPRAppl23}
%%%%
\begin{equation}
    \label{eq:polariton_freq}
    \Omega_\pm = \frac{1}{\sqrt{2}} \sqrt{\omega_{c}^2 + \omega_{b}^2 \pm \sqrt{\left(\omega_{c}^2 - \omega_{b}^2 \right)^2 + 16 \omega_{c} \omega_{b} g^2}} ~,
\end{equation}
where $\omega_b$ is correlated to both magnetic field and temperature assuming that
%%%%
\begin{equation}
    \label{eq:spin-wave-Meissner-current_freq}
    \omega_b=\omega_{0}+\delta_{sc}, 
\end{equation}
%%%%%%%
where $\omega_{0}$ the frequency of the lowest YIG mode (Eq.~\ref{eq:KS_film_0}) and $\delta_{sc}$ a temperature-dependent change. The latter can be quantified by self-consistently including the spin-wave induced Meissner currents in the Landau-Lifshitz-Gilbert equation, to obtain \cite{BorstScience23}
\begin{equation}
    \delta_{sc} \approx \gamma \mu_0 M_s k_y d r \frac{1-e^{-\frac{2t}{\lambda_L}}}{(k_y \lambda_L+1)^2-(k_y \lambda_L-1)^2 e^{-\frac{2t}{\lambda_L}}},
    \label{eq:delta}
\end{equation} 
where $t$ is the thickness of the YBCO film and $r$ is a dimensionless geometrical factor associated with the YIG thickness and spin-wave ellipticity. $\lambda_L$ is the penetration depth of YBCO, which in the simplest two-fluid binomial approximation reads \cite{ProzorovSuperScieTech06}
%%%%
\begin{eqnarray}
   \lambda_L=\frac{\lambda_L(0)}{\sqrt{1-\left (\frac{T}{T_c} \right)^{p}}},
   \label{eq:penetration_depth}
\end{eqnarray}
%%%%
where $T_c$ is the critical temperature, $\lambda_L(0)$, is the London penetration length in the zero temperature limit, and $p=4/3$ for d-wave superconductors \cite{ProzorovSuperScieTech06, VendikIEEE98, GhigoSuperSciTech04}.

%%%
\begin{figure}[t]
\centering
\includegraphics[width=8cm]{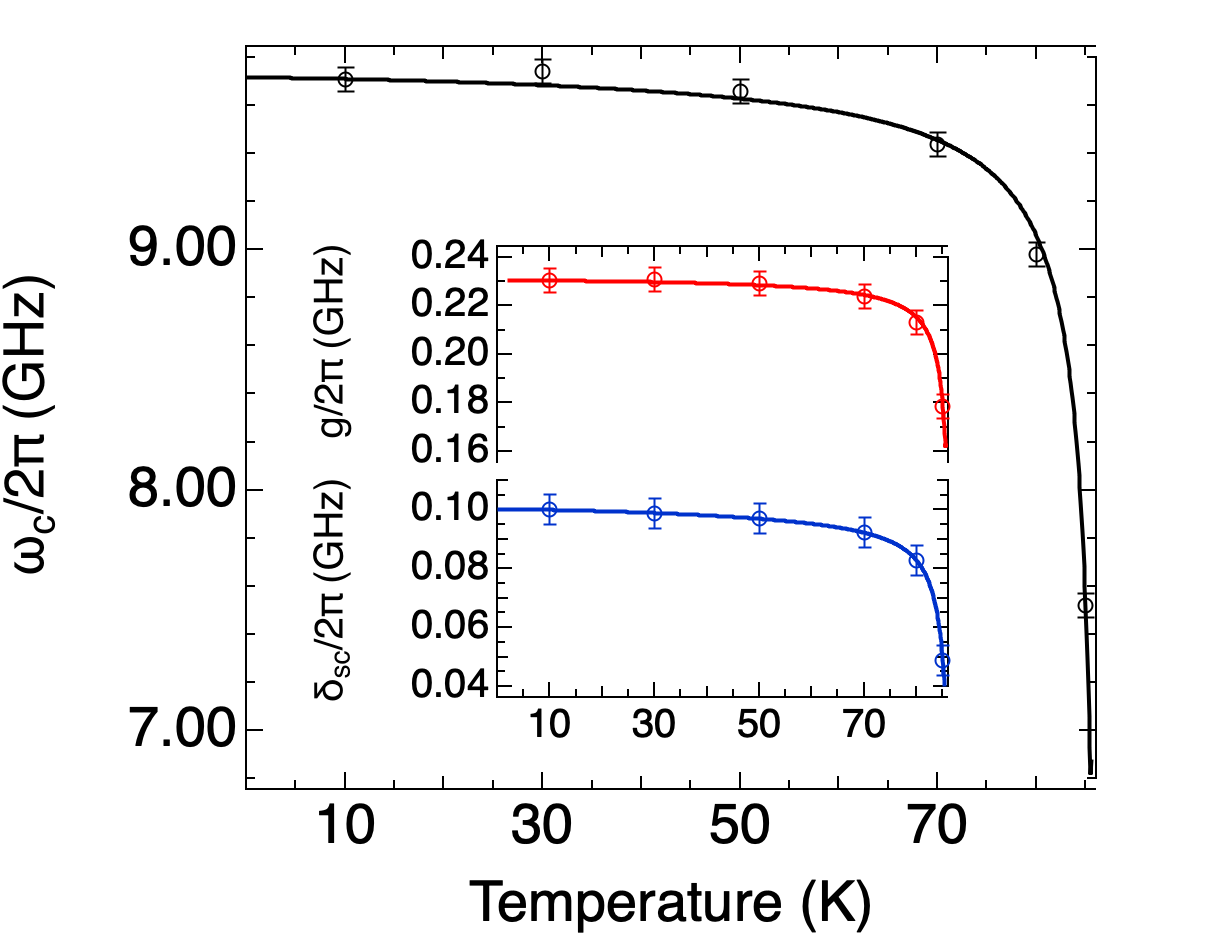}
\caption{Frequency of the resonator extracted from transmission maps acquired at different temperatures (circles). The solid line shows the fit with Eq.~\ref{eq:omega0_temp}. Inset. Coupling strength and frequency shift calculated using, respectively, the resonator frequencies (circles) and Eq.~\ref{eq:omega0_temp} (solid lines). Error bars are estimated from the uncertainty of the fit parameters.}
\label{fig:resonator_param}
\end{figure}
%%%

By means of Eq.~\ref{eq:polariton_freq} we have fitted the evolution of polaritons in Fig.~\ref{fig:resonator_data}. The frequency of the resonator, $\omega_c (T)$, is used at each temperature as a free parameter; $\omega_b$ has been calculated using Eq.~\ref{eq:delta}, where $r=0.45$ %NB it was 0.113-0.122 for the 5 um film
and the penetration depth has been derived from Eq.~\ref{eq:penetration_depth} with $T_c=86.1~\mathrm{K}$ and $\lambda_L(0)=97~\mathrm{nm}$. The latter parameters are obtained from the fit of $\omega_c(T)$ (Fig.~\ref{fig:resonator_param}) using \cite{GhigoSuperSciTech04}
%%%%
\begin{equation}
\omega_c \approx \omega_c(0) \sqrt{\frac{L[\lambda_L(T_0)]}{L[\lambda_L(T)]}},
\label{eq:omega0_temp}
\end{equation}
%%%%
being $L[\lambda_L(T)]$ the temperature-dependent inductance and $\omega_c(0)/2\pi=9.7~\mathrm{GHz}$ at $T_0=10~\mathrm{K}$. Finally, the collective coupling $g=g_{s} \sqrt{2 s_{\mathrm{Fe}} N_s}$, where $s_{\mathrm{Fe}}=5/2$ is the single-ion spin of Fe$^{3+}$, is calculated assuming that the number of spins is $N_s=1.49 \times 10^{13}$ in the entire temperature range. The temperature dependence of $g$ derives, through $\omega_c(T)$, from the spin-photon coupling
%%%%%
\begin{equation}
    g_{s} \approx \frac{\mu_0 \omega_{c}}{4 w} \sqrt{\frac{h}{Z_0}},
\label{eq:gs}
\end{equation}
%%%%
where $b_{vac}$ is the vacuum magnetic field of the resonator and $Z_0=58~\mathrm{\Omega}$ is the nominal impedance of the CPW line \cite{GhirriPRAppl24}. 

Fig.~\ref{fig:resonator_param} shows the temperature evolution of the parameters that describe the experimental data. From the fitted values of $\omega_c(T)$ and using the values of $\delta_{sc}(T)$ and $g(T)$ calculated with $r=0.45$ and $N_s=1.49 \times 10^{13}$, we can reproduce the evolution of the transmission spectra shown in Fig.~\ref{fig:resonator_data} for temperatures between 10 and 85~K.

\section{Discussion}
\label{sec:discussion}

It is worth comparing the results obtained here for the 104-nm-thick YIG film with those reported for thicker films \cite{GhirriPRAppl23, GhirriPRAppl24}. The broadband spectra shown in Figs.~\ref{fig:broadband_1} and \ref{fig:broadband_2} display two well-separated resonances at frequencies $\omega_0$ and $\omega_1$, in contrast with the larger number of closely spaced lines observed for a 5-$\mathrm{\mu m}$-thick YIG film under similar experimental conditions \cite{GhirriPRAppl23, GhirriPRAppl24}. Their dependence as a function of $H_0$ has been analyzed with Eqs.~\ref{eq:KS_film_0} and \ref{eq:KS_film_1}, using fitting parameters compatible with those previously reported \cite{MaierFlaigPRB17, ZyuzinJETPLett96, GhirriPRAppl23, GhirriPRAppl24}. In particular, Eq.~\ref{eq:KS_film_1} shows, in accordance with the trend observed in the experiments, an increase in the frequency $\omega_1$ when the thickness of the YIG film decreases. The chosen fitting parameters account for the non-trivial evolution of the FMR line at low temperature, which has recently been attributed to the increase in magnetic anisotropy as a result of the interaction between the YIG film and the GGG substrate \cite{SerhaNPJSpin24, SchmollPRB25, KimPRMater25}. We note that the temperature evolution of the resonance position in Figs.~\ref{fig:broadband_1}(c) and \ref{fig:broadband_2}, and the deviation from the Lorentzian line shape in the resonance spectra, are in line with other recently reported experimental \cite{SerhaNPJSpin24, KimPRMater25}.

The broadening of the lowest resonant peak and the diminishing of its amplitude at low temperature evidence the increase of losses (Fig.~\ref{fig:broadband_1}(c)). Despite the narrow FMR linewidths obtained at room temperature \cite{DubsJPhysDApplPhys17, RaoJPhysD18}, low-temperature measurements on YIG/GGG films have shown FMR linewidths up to 30~MHz due to local defects and other effects induced by the GGG substrate \cite{SerhaNPJSpin24, SchmollPRB25, KimPRMater25}. Moreover, experiments with YIG films having conductive electrodes on top have shown the broadening of the FMR line \cite{RezendeAPL13, BunyaevPRAppl20, SchmollArxiv25}. Additional contributions to line broadening come from the superconductor. Inhomogeneities in the local magnetic field can be related to Meissner screening and non-perfect parallelism of the film surface with respect to the external magnetic field. Being our YBCO film a type-II superconductor with pinned vortices, in the mixed state we also expect the presence of inhomogeneous magnetic fields at the interface with the YIG film due to the presence of the Abrikosov lattice.

Additional fingerprints of the interplay between magnetic and superconducting layers can be found in the temperature-dependent properties of the YBCO CPWs. The amplitude of the microwave field decreases quasiexponentially as the height from the CPW increases \cite{GhirriPRAppl23}, additionally, at low $T$ the penetration depth of YBCO drops to $\approx150$~nm \cite{GhirriPRAppl24}. The appearance of higher magnon modes in broadband measurements (Fig.~\ref{fig:broadband_2}) is likely due to the coupling between tightly confined microwaves and stray fields \cite{MaksymovPhysE15}. Furthermore, experiments with thick YIG films and YBCO resonators in the ultrastrong coupling regime evidenced the shift of the anticrossing position with respect to the unperturbed resonance line of the YIG film ($\delta_{sc}/2\pi$ up to $1.2~\mathrm{GHz}$) \cite{GhirriPRAppl24}. This behavior has been interpreted as an effect of electromagnetic proximity between the magnetic and superconducting layers \cite{BorstScience23, YuArxiv25}. 

In the present case, the 104-nm-thick YIG film gives rise to a much smaller shift ($\delta_{sc}/2\pi =0.1~\mathrm{GHz}$), which is justified by the smaller $d$ in Eq.~\ref{eq:delta}. However, the broadband spectra in Fig.~\ref{fig:broadband_2} suggest that the line position at different temperatures can be reproduced using only the parameters of the YIG, without including the effect of the superconductor. We note that in Eq.~\ref{eq:delta} the geometric parameter $r$ has been used as a free fitting parameter to adjust the absolute value of $\delta_{sc}$; to better quantify its magnitude, a more systematic study that involves variations in the size and thickness of the YIG film would be required. Furthermore, we note that the estimated magnon and photon numbers are $N_m=N_p \approx 10^{10}\ll N_s$ \cite{GhirriPRAppl23}. The shifts in magnon frequency extracted from Fig.~\ref{fig:resonator_data} are greater than those expected from the magnon Kerr effect in our experiment \cite{WangPRB16, ZhangSciChina19}. Considering the estimated mode volume of the resonator $V_m=N_s/\rho = 7 \times 10^{-14}~\mathrm{m^3}$, being $\rho=2.1 \times 10^{28}~\mathrm{m^{-3}}$ the spin density of the YIG, and the first-order anisotropy constant of the YIG film ($K_{an}$), we obtain the Kerr coefficient $K=\hbar\mu_0 K_{an}\gamma^2/(M_s^2 V_m) \approx 10^{-7}~\mathrm{Hz}$. The Kerr effect results much lower than the collective spin-photon coupling. These numbers further confirm our analysis carried out in the linear regime with a low number of excitations.

We finally make a direct comparison between the coupling strength obtained and those reported in \cite{GhirriPRAppl23, GhirriPRAppl24} using thicker YIG/GGG films and identical YBCO CPW resonator under similar experimental conditions. The maximum coupling strength derived from the spectra in Fig.~\ref{fig:resonator_data} is $0.23~\mathrm{GHz}$ while the coupling obtained with a 5-$\mathrm{\mu m}$-thick YIG film having similar lateral size is around $1.1~\mathrm{GHz}$ \cite{GhirriPRAppl24}. The ratio between thicknesses is 50 while the coupling strength achieved with the thicker film is only 5 times higher. By using wider 5-$\mathrm{\mu m}$-thick films, the coupling strength increased to around 2~GHz, although no significant improvement was observed for thicknesses as high as 20~$\mathrm{\mu m}$ \cite{GhirriPRAppl23}. These results confirm that in our experiments a significant coupling with the CPW is achieved at distances below $1~\mathrm{\mu m}$ above the YBCO surface.

\section{Conclusions}
\label{sec:conclusions}

In conclusion, we have studied the microwave response of nonconductive 104-nm-thick YIG films positioned on top of a YBCO superconducting layer. Transmission spectra were acquired at different temperatures either by using a broadband CPW or a CPW resonator. In the former case, experimental data have been essentially reproduced using well-consolidated models and physical quantities describing the spin-wave excitation spectrum of the YIG film. In the latter case, the coupling extracted from the splitting of magnon-photon polaritons amounts to 230~MHz. The changes in the polaritonic spectrum at different temperatures can be attributed to the effects of the penetration depth in the YBCO resonator. The comparison between results obtained with YIG films having different thicknesses suggests a non-trivial role of the system geometry and material parameters in determining the evolution of the transmission spectra.

\section*{Acknowledgments}

This work was partially supported by the project SMILE-SQUIP funded by the Italian NQSTI - National Quantum Science and Technology Institute code PE0000 0023 and by the U.S. Office of Naval Research Award No. N62909-23-1-2079. A.G. acknowledges financial support from PNRR MUR project ECS\_00000033\_ECOSISTER.

%% If you have bib database file and want bibtex to generate the
%% bibitems, please use
%%
%\bibliographystyle{elsarticle-num} 
%\bibliography{biblio}

%% else use the following coding to input the bibitems directly in the
%% TeX file.

%% Refer following link for more details about bibliography and citations.
%% https://en.wikibooks.org/wiki/LaTeX/Bibliography_Management

%\begin{thebibliography}{00}

%% For numbered reference style
%% \bibitem{label}
%% Text of bibliographic item

%\bibitem{lamport94}
  %Leslie Lamport,
  %\textit{\LaTeX: a document preparation system},
  %Addison Wesley, Massachusetts,
  %2nd edition,
  %1994.

%\end{thebibliography}

\end{document}